\documentclass[journal=jacsat,manuscript=article]{achemso}
\usepackage{chemformula} 
\usepackage[T1]{fontenc} 
\usepackage[colorinlistoftodos]{todonotes}
\usepackage{amsmath}
\usepackage{amssymb}
\usepackage{mhchem}
\usepackage[export]{adjustbox}
\usepackage{graphicx}
\usepackage{subcaption}
\usepackage[colorinlistoftodos]{todonotes}
\usepackage[colorlinks=true, allcolors=blue]{hyperref}
\usepackage{booktabs}
\usepackage{tabularx}
\usepackage{listings}
\usepackage{xcolor}
\usepackage{float}
\usepackage{chngpage}
\usepackage{longtable}

\author{Chaoyi Chang}
\author{Andrew J. Medford}
\email{ajm@gatech.edu}
\phone{}
\fax{}
\affiliation[Georgia Institute of Technology]
{School of Chemical \& Biomolecular Engineering, Georgia Institute of Technology}

\title[An \textsf{achemso} demo]
  {Application of DFTB and machine learning to evaluate the stability of biomass intermediates on the Rh(111) surface}
   
\begin{document}
\begin{abstract}
Biomass compounds adsorbed on surfaces are challenging to study due to the large number of possible species and adsorption geometries. In this work, possible intermediates of erythrose, glyceraldehyde, glycerol and propionic acid are studied on the Rh(111) surface. The intermediates and elementary reactions are generated from first 2 recursions of a recursive bond-breaking algorithm. These structures are used as the input of an unsupervised Mol2Vec algorithm to generate vector descriptors. A data-driven scheme to classify the reactions is developed and adsorption energies are predicted. The lowest mean absolute error (MAE) of our prediction on adsorption energies is 0.39 eV, and the relative ordering of different surface adsorption geometries is relatively accurate. We show that combining geometries from density functional tight-binding (DFTB) calculations with energies from machine-learning predictions provides a novel workflow for rapidly assessing the stability of various molecular geometries on the Rh(111) surface. 
\end{abstract}

Biomass compounds are playing a key role in biorenewable products and are the basis of a sustainable economy. For example, the US Department of Energy has listed the 10 top biorefinery products with the highest potential impact \cite{werpy2004top}. Selective conversion between those products and other small molecule products are important if biomass compounds are to form the basis of a sustainable economy. Rh is a commonly-used transition metal catalyst for conversion of biomass compounds and has been the subject of numerous experimental studies. For example, Rh catalysts have been used for succinic acid conversion to fumaric acid \cite{lam2014carbon}, hydrogenolysis of furfural to 1,2-pentanediol \cite{pisal2019single} and production of C1 compounds from ethanol \cite{zhang2014ethanol, cavallaro2000ethanol, mei2016steam}. However, there have been relatively few systematic computational studies of biomass intermediates on Rh surfaces \cite{abdelfatah2019prediction}. Density functional theory (DFT) is the most common theory used to study the adsorption and reaction of biomass compounds \cite{auneau2011unravelling, yang2020investigation, wang2013glycerol}. However, complex molecules have multiple binding sites and various geometries, and DFT calculations of larger molecules are expensive and require significant computational effort to converge. Thus, previous studies often utilize empirical and machine learning (ML) methods to relate reaction properties of biomass molecules or intermediates to their structures, physical properties and even experimental conditions \cite{goldsmith2018machine, gu2018thermochemistry, yang2020investigation, vorotnikov2014group}. Most of these ML methods could reach a mean absolute error of adsorption energies/transition state energies within 0.4 eV, \cite{vorotnikov2014group, salciccioli2012adsorption, abdelfatah2019prediction}  with some examples of predictions as accurate as 0.2 eV using a combination of physical and structural features along with feature selection methods \cite{chowdhury2018prediction, gu2018thermochemistry}.

As mentioned previously, a key challenge with biomass molecules is their complexity, and ``model compounds'' are often used to simplify systems of interest. A ``model compound'' refers to a relatively small molecule (typically fewer than 8 heavy atoms) that can be used for studying a larger compounds with similar chemical properties and functional groups. Previous experimental studies showed that the reaction pathway of \ce{glyceraldehyde <=> glycerol} is similar to \ce{linear glucose <=> sorbitol} \cite{auneau2011unravelling,valter2020partial,wang2013glycerol,kwon2013electrocatalytic} and that propionic acid ketonization is similar to larger carboxylic acid ketonization on Rh surfaces \cite{yang2020investigation,kumar2018ketonization}. While model compounds are commonly used, they are typically identified by heuristics and intuition. However, a data-driven strategy for systematic identification of model compounds was proposed in a prior study by the authors. \cite{chang2020classification} It was pointed out that glycerol, glyceraldehyde, propionic acid, and erythrose are good model compounds to study larger molecules like sorbitol and linear-structured glucose. This makes these four molecules a key starting point for computational and experimental studies seeking a more general understanding of the catalytic conversion of biomass derivatives on solid surfaces. 

In this study, we extend the previously-developed embedding models of gas-phase formation energies \cite{chang2020classification} to adsorbed surface intermediates on the Rh(111) surface. A total of 171 intermediates from the first 2 bond-breaking recursions of erythrose, glyceraldehyde, glycerol and propionic acid are studied. Mol2Vec is used for generating vector descriptors and 83 clusters based on 6 reaction types (C-C, C-O, O-H, C-H, C-M, O-M) are obtained from single-group ``radius zero'' (R0) Mol2Vec descriptors. Linear discriminant analysis (LDA) and partial least squares (PLS) are used for dimensional reduction of two- and three-group ``radius one'' (R1) Mol2Vec descriptors, providing low-dimensional vector descriptors for each adsorbate. These vectors are combined with a linear least-squares regression model, yielding a mean absolute errors (MAE) as low as 0.39 eV. Finally, pre-optimization via density-functional tight binding (DFTB) is combined with our embedding models to establish a workflow for rapidly identifying stable adsorption geometries. We show that this workflow identifies 20 new lowest-energy geometries for 171 adsorbates studied, indicating that systematic approaches for identifying the lowest-energy structures of large adsorbed molecules are a necessary addition to the tool set of computational catalysis.

\begin{figure}[h]
 \centering
 \includegraphics[width=\linewidth]{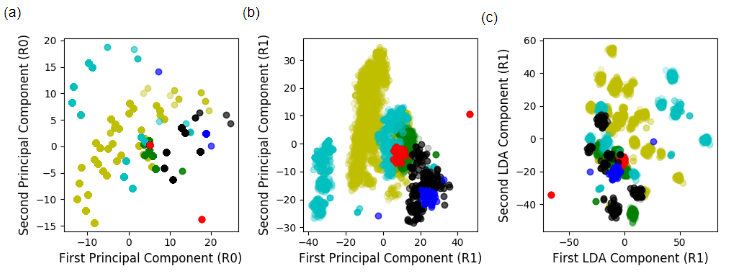}
\caption{Visualization of reaction vector clusters and 6 reaction types (yellow for C-C, green for C-H, cyan for C-O, red for O-H, black for C-M, blue for O-M): (a) 1st vs. 2nd component of PCA (R0), (b) 1st vs. 2nd component of PCA (R1) and (c) 1st vs. 2nd component of LDA (R1 with class labels from R0 clusters).}
\label{fig:PCA_LDA}
\end{figure}

First, we establish an approach to assign detailed classes for each bond type of adsorbed biomass intermediates, similar to the previously-developed approach for gas-phase molecules \cite{chang2020classification}. Mol2Vec \cite{jaeger2018mol2vec} is used for generating vector descriptors for intermediates and reactions with 200 dimensions in R0 and R1 \cite{chang2020classification}. The traditional extended-connectivity fingerprint (ECFP) is modified to include surrounding heavy atom number, surrounding hydrogen atom number, valence, electric negativity and mass as invariants for atoms contained in a structure. Metal atoms are still considered to be heavy atoms but all the other properties are considered as NaN so that metal atoms are a general rather than specific type (e.g. Rh is indistinguishable from any other metal). An algorithm (see SI) to add metal atoms to unsaturated C, O atoms one-at-a-time is applied to the 171 intermediates generated from the first 2 bond-breaking recursions of erythrose, glyceraldehyde, glycerol and propionic acid, and additional structures are generated by DFTB minima hopping calculations (see below), yielding a total of 2,498 adsorbed structures. The adsorbed structures are combined with 91,098 gas-phase species, resulting in a total of 93,569 structures that are used as the corpus for training the Mol2Vec model (see SI for details). We also utilize 6 basic types of reactions for visualization and checking intuition. These 6 types include 4 types of intra-adsorbate reactions, C-C (C=C and C-C), C-H, C-O (C=O and C-O), OH, and 2 types of elementary reactions with metal atoms, C-M (C-Metal), O-M (O-Metal). The single and double bond breaking reactions are considered as a single class since the distinction between them becomes ambiguous for adsorbed species due to partial bond orders and conjugation.  A total of 13,422 gas-phase reactions from our previous work \cite{chang2020classification} and 1,666 additional elementary surface reaction steps are included for the analysis of reactions. For the convenience of visualization, principal component analysis (PCA) is used for reducing the 200-dimensional reaction vectors to 2 dimensions, and each type of reaction is represented by a different color (yellow for C-C, green for C-H, cyan for C-O, red for O-H, black for C-M, blue for O-M). Fig. \ref{fig:PCA_LDA}a and \ref{fig:PCA_LDA}b show the visualization of R0 and R1 reaction vectors for the full corpus of intermediates and reactions.

The PCA result in R0 indicates that there are discrete well-defined clusters within the 15,088 reactions. The Euclidean distance between reactions of the original 200-dim R0 vectors are calculated to identify distinct clusters, and the cutoff to separate clusters is set to be 0.05. A total of 83 clusters in R0 are obtained. Each of the 83 clusters contain reactions with different bond-breaking types and different atomic environments of the the atoms within the elementary reaction. The atomic environment refers to the surrounding heavy atoms and hydrogen atoms of the 2 reacting atoms. Table S1 shows all details of the atomic environments for each of these reaction types.  

The obtained 83 R0 clusters are then used as labels for a supervised classification on R1. Linear discriminant analysis (LDA), is applied on the 200-dimension R1 vector descriptors together with the 83 classes as class labels. Fig. \ref{fig:PCA_LDA}c shows the first and the second LDA components of LDA-projected R1 reaction vectors, and the 6 reaction types are represented by 6 different colors for the convenience of visualization. The R1 vectors are reduced to 1-82 dimensions via the LDA projection. The LDA projections represent an unsupervised vector space for analyzing reactions and intermediates, and are used to predict adsorption energies on Rh (111) surface.

Ultimately, the goal is to predict adsorption energies of biomass species. The 171 adsorbates, ranging from C1 - C4 species (see SI for details) included in this study are generated from the first two bond-breaking recursions \cite{chang2020classification} of erythrose, glyceraldehyde, glycerol and propionic acid. In general, each species can have multiple adsorption energies due to differences in molecular configuration and binding sites. This makes it challenging to directly predict the most stable binding site even with DFT. For this reason, a low-cost pre-optimization tool is needed to identify different stable geometries. Density functional tight-binding (DFTB) provides a rapid physics-based route that provides relatively accurate energies and geometries of reactive surfaces \cite{porezag1995construction, seifert1996calculations}. We use the open-source Hotbit \cite{koskinen_CMS_09} Python package with a previously-developed parameterization for Rh/C/H/O \cite{yang2016intrinsic} for performing DFTB. The Hotbit calculator is used with  constrained minima hopping \cite{peterson2014global} where Rh slab atoms have fixed positions and adsorbate bonds have fixed lengths \cite{ase-paper} to generate local minima geometries for each adsorbate. 

The DFTB-based minima hopping process is used to generate up to 50 adsorption geometries for each of the 171 adsorbates, yielding a total of 857 different adsorbate structures. Geometries that are within 1 eV of the lowest Hotbit energies are then calculated by DFT. DFT calculations are performed using Quantum ESPRESSO \cite{giannozzi2009quantum} with the PBE\cite{perdew1996generalized} exchange-correlation functional at a planewave cutoff of 450 eV (see SI for details). This process yields 328 unique geometries and associated adsorption energies, which are used as inputs for supervised training of the regression models.

\begin{figure}[h]
    \makebox[\linewidth][c]{%
    \adjustbox{minipage=1.3em,valign=t}{\subcaption{}\label{fig:hb_dist}}%
  \begin{subfigure}[t]{\dimexpr.5\linewidth-1.3em\relax}
  \centering
  \includegraphics[width=\linewidth,valign=t]{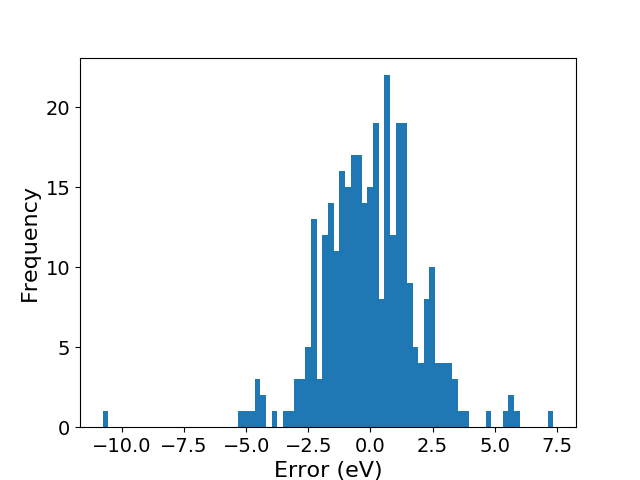}
  \end{subfigure}%
  \adjustbox{minipage=1.3em,valign=t}{\subcaption{}\label{fig:dft_hb_lr}}
  \begin{subfigure}[t]{\dimexpr.5\linewidth-1.3em\relax}
  \centering
  \includegraphics[width=\linewidth,valign=t]{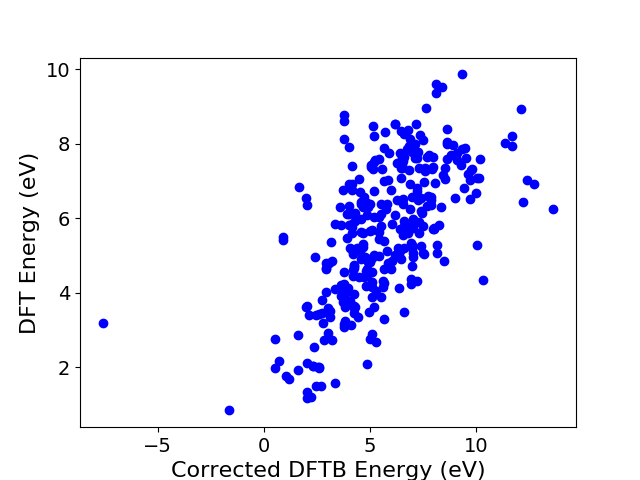}
  \end{subfigure}
    }
\caption{Accuracy of DFTB energies after reference alignment (a) error distribution and (b) parity plot of DFTB vs. DFT energy}
\label{fig:hb_results}
\end{figure}

\begin{figure}[!ht]
\makebox[\linewidth][c]{%
\adjustbox{minipage=1.3em,valign=t}{\subcaption{}\label{fig:regression}}
  \begin{subfigure}[t]{\dimexpr.5\linewidth-1.3em\relax}
  \centering
  \includegraphics[width=\linewidth,valign=t]{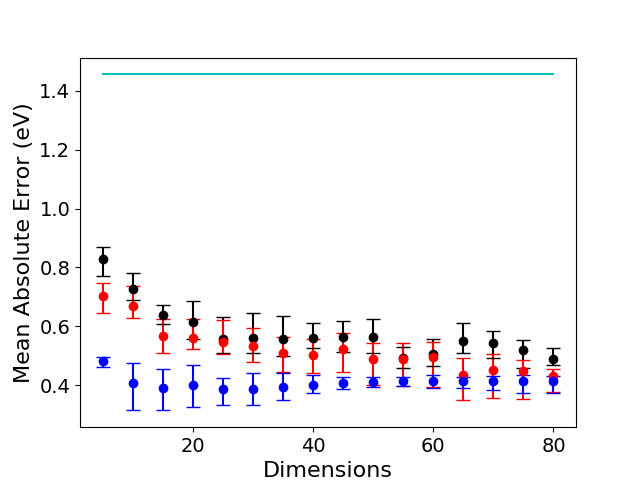}
  \end{subfigure}%
    \adjustbox{minipage=1.3em,valign=t}{\subcaption{}\label{fig:spearman_hb}}%
  \begin{subfigure}[t]{\dimexpr.5\linewidth-1.3em\relax}
  \centering
  \includegraphics[width=\linewidth,valign=t]{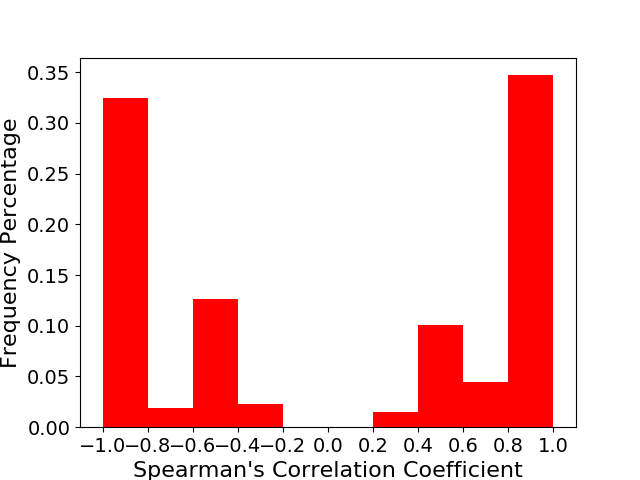}
  \end{subfigure}
    }\\
\makebox[\linewidth][c]{%
    \adjustbox{minipage=1.3em,valign=t}{\subcaption{}\label{fig:spearman_pls}}%
  \begin{subfigure}[t]{\dimexpr.5\linewidth-1.3em\relax}
  \centering
  \includegraphics[width=\linewidth,valign=t]{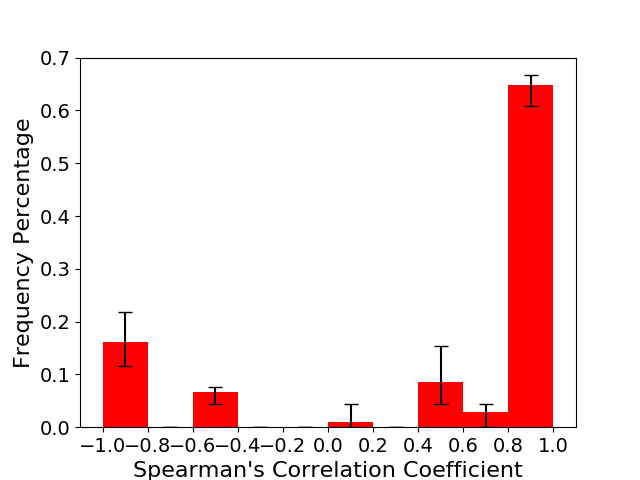}
  \end{subfigure}%
  \adjustbox{minipage=1.3em,valign=t}{\subcaption{}\label{fig:spearman_lda}}
  \begin{subfigure}[t]{\dimexpr.5\linewidth-1.3em\relax}
  \centering
  \includegraphics[width=\linewidth,valign=t]{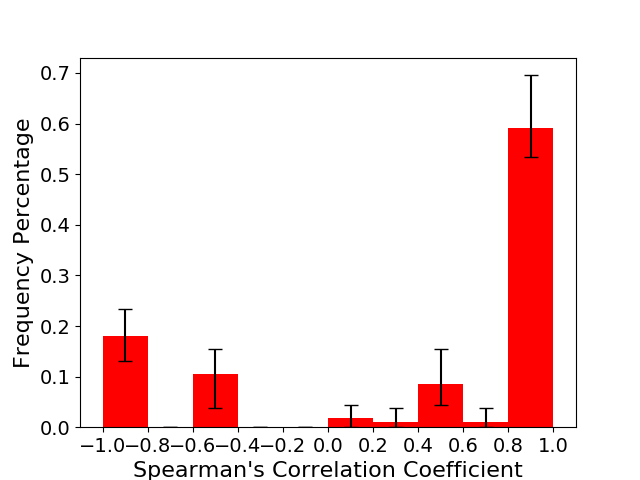}
  \end{subfigure}
    }
\caption{DFTB and ML results of (a) cross validation error of OLS (black), LDA (red), PLS (blue) and DFTB MAE (cyan line), spearman's correlation coefficient of (b) DFTB calculations, (c) 30-dim PLS regression and (d) 65-dim LDA projections with error bars}
\label{fig:spearman}
\end{figure}

Adsorption energies with both DFT and DFTB are computed for all 328 unique geometries. This data is used to assess the accuracy of the DFTB energies. Since DFT and DFTB do not use a common reference, it is necessary to align the energies based on the stochiometry of each adsorbate:
\begin{equation}
    E_{DFT}=E_{DFTB}+\left( \sum_{i \in [C,H,O]} c_{i}*n_{i} \right)+\epsilon
\end{equation}
where $E_{DFT}$ is the DFT adsorption energy, $E_{DFTB}$ is the DFTB adsorption energy, $n_{i}$ is the number of C, H, O atoms in the adsorbate, $c_{i}$ are fitted coefficients and $\epsilon$ is the residual error. Fig. \ref{fig:hb_dist} shows the distribution of the residual DFTB error in energy calculation and Fig. \ref{fig:dft_hb_lr} shows the parity plot of DFT and corrected DFTB energies. The mean absolute error (MAE) of DFTB is 1.46 eV, with notable outliers that can have errors of $>$5 eV. In addition, we use the Spearman's correlation coefficient to evaluate the ability of DFTB to correctly order the energies of different geometries for a given adsorbate. The results, shown in Fig. \ref{fig:spearman_hb}, reveal that DFTB yields incorrect ordering of adsorbates more than 60\% of the time. Despite these large energy errors, the geometries are more accurate with an average position difference of 1.04 \AA{}. This suggests that while Hotbit is a reasonable tool for generating adsorbate geometries, the energy predictions are not sufficiently accurate to yield chemical insight or even correctly order the stability of various adsorbate geometries.

To improve the accuracy of energy predictions we turn to a supervised ML approach. The workflow utilizes the Mol2Vec vectors for each adsorbate as inputs, similar to  our previously-developed approach for gas-phase energies \cite{chang2020classification}. We compare three different linear models for predicting adsorption energies from Mol2Vec vectors: The original unsupervised Mol2Vec vectors (OLS), the semi-supervised LDA projection, and the supervised partial least squares (PLS) projection. In each case we utilize a 75/25 train/test split with 4 random repeats with the results from the 328 DFT calculations as the target. The \texttt{scikit-learn} package \cite{scikit-learn} is used for each algorithm (see SI for more details). In each case we evaluate the mean absolute error of the test set as a function of the dimension of the input vector, with a maximum dimension of 82 (the maximum dimension of the LDA vectors). 

The results of the ML predictions, shown in Fig. \ref{fig:regression} show that the lowest MAE of OLS, PLS and LDA are $\sim$0.51 eV, $\sim$0.39 eV and $\sim$0.43 eV at $\sim$55-dim, $\sim$30-dim and $\sim$65-dim respectively. The MAE of OLS, PLS and LDA decrease at first and reach a plateau or increase slowly after 55-dim, 30-dim and 65-dim. The lowest MAE of the LDA and PLS models are very similar, and we expect that the LDA model will be more transferrable to other adsorbates since the adsorption energies are not used to generate the feature vector. By contrast, we expect that the PLS performance is more specific to this set of adsorbates since the adsorption energies are used to generate the inputs.  Comparing the results of the machine-learning models to DFTB, we see that the prediction errors are much lower ($\sim$0.4 eV for ML vs. $\sim$1.5 eV for DFTB). While the error of $\sim$0.4 eV is still somewhat larger than the typical DFT error ($\sim$0.2 eV), the ordering of energies of different geometries with the ML model is relatively good, with a correct ordering at least 59\% of the time (Fig. \ref{fig:spearman_pls} and Fig. \ref{fig:spearman_lda}). However, the ML models requires geometries as inputs, which must be generated by DFTB. This suggests a synergistic approach between the models is required, where DFTB is used to identify geometries and ML is used to predict energies.


The average error of the best machine-learning models ($\sim$0.4 eV) is relatively large, indicating that DFT will need to be used when accurate energies are required. However, the number of possible geometries and active sites for biomass molecules makes brute force DFT calculations impractical for large numbers of adsorbates. The combination of geometries from DFTB along with predictions from ML can alleviate this issue by identifying the geometries that are most likely to be stable, thus reducing the number of DFT calculations required. Spearman's correlation coefficient is used to quantitatively assess the model's ability to predict the energy order for different geometries of the same adsorbate. We utilize a 75/25 train/test split with 4 random repeats with the adsorbates for PLS and LDA (approximately 1-3 geometries per adsorbate). Fig. \ref{fig:spearman_pls} and Fig. \ref{fig:spearman_lda} show the distribution of Spearman's correlation coefficient of PLS and LDA test sets with error bars. PLS and LDA obtain more than 65\% and 59\% correct energy orders on average with no more than 25\% totally inverted (88\% of the totally inverted geometries include only 2 geometries). Fig. \ref{fig:spearman_hb} shows the distribution of Spearman's correlation coefficient of DFTB with only 35\% correct, while more than 30\% are totally inverted.

This demonstrates that both PLS and LDA have similar performance, but both are better than DFTB in predicting energy orders. However, there are still some mis-ordered energies based on the ML predictions, suggesting that the lowest energy structure may not be correctly identified in all cases. To increase robustness, we utilize the average standard deviation of the model errors (0.45 eV) as a tolerance factor, meaning any structure within 0.45 eV of the lowest energy is considered as a possible global minimum. Using simple estimates from probability theory this corresponds to 75\% confidence that the true global minimum will be included (see SI). This cutoff can be adjusted to improve confidence at the expense of more DFT calculations. Both PLS and LDA models are built based on the 328 DFT calculations and are used for predicting all 857 geometries generated from DFTB and minima hopping. The minimum energy of each adsorbate and the geometries within 0.45 eV of the minimum energy are extracted, and any geometries that have not already been computed are calculated with DFT. Both of the ML models are used, and the geometries are calculated with DFT if the energy predicted by either model is within the threshold.

\begin{figure}[h]
    \centering
    \includegraphics[width=\linewidth]{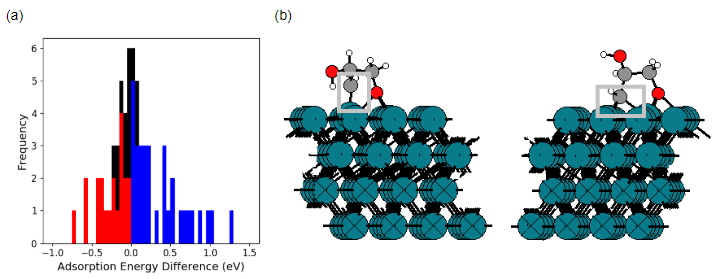}
    \caption{Results of DFT calculations for structures predicted to be low-energy by ML model. (a) Stacked bar plot of adsorption energy difference for structures with the same (black) or different (red/blue) SMILES strings from prior global minimum, with lower-energy structures in red and higher-energy structures in blue. (b) A representative example of an adsorbate ($[O]CC(O)[CH]$) where the new lower-energy geometry binds with a qualitatively different structure, where $[CH]$ binds directly to a single Rh atom (left, gray box) instead of two Rh atoms in the previous structure (right, gray box).}
  \label{fig:diff_example}
\end{figure}

The ML model identifies multiple possible new global minima for 65 of the adsorbates, corresponding to 154 additional DFT calculations. The MAE between the model predictions and the DFT energies of the new structures are 0.60 eV and 0.62 eV for LDA and PLS models, about 50\% higher than the MAE of the test set. The results of these calculations reveal that the energies of many of these structures are lower than the previously-determined lowest energy structure, as shown in Fig. \ref{fig:diff_example}a (see SI for details). For 20 of the 65 adsorbates the configuration of the global minima was sufficiently different to lead to a new SMILES representation.  For example, for the the adsorbate $[O]CC(O)[CH]$ the ML model identifies a $[CH]$ bidentate-binding geometry with 0.44 eV lower energy than the monodentate geometry that was previously identified as the lowest energy structure, as shown in Fig. \ref{fig:diff_example}b. For 13 other adsorbates, the difference in geometry was less drastic so that the SMILES string of the new structure was the same as the old structure, but the energy was slightly different by $<$0.23 eV (see Fig. \ref{fig:diff_example}b). The original energy was lower than all newly-computed energies in 32 adsorbates. Overall, the original workflow for identifying global minima failed to identify the global minimum for at least 17\% of adsorbates, and had qualitative differences in binding geometries (different SMILES string or energy difference $>$ 0.05 eV) for 14\% of adsorbates.

The results of this work suggest that a combination of physical approximations and ML models is a promising route toward identifying global minima of complex adsorbates. A general workflow involves a first step that uses an approximate physical method (DFTB in this case) to rapidly generate many candidate geometries. The second step involves using DFT to calculate the energies of the most stable structures (structures within 1 eV of the minimum in this case, leading to 328 geometries of 171 adsorbates). Third, these DFT energies are used to train ML models, here based on Mol2Vec and linear regression, and the ML models are then used to predict the energies of all candidate structures. Finally, the predictions of the ML model are used to identify new structures that will be computed with DFT, in this case structures within 0.45 eV (the standard error of the ML models on the test set) of the predicted minimum. The results of this work show that this process yields 20 global energy minima that would have been incorrect without the use of the ML model. This process can be made more efficient with improved physical approximations and more accurate ML models, which may be necessary to tackle larger and more complex biomass molecules.

The size and complexity of biomass molecules leads to a major challenge in predicting the global minimum adsorption geometry, and this challenge is compounded by the number of possible intermediates that appear in biomass reaction networks. It is clear that new techniques are needed to accelerate the study of these systems since direct calculation with DFT or other quantum chemical techniques is impractical. Here, we show that ML models based on Mol2Vec descriptors can achieve an MAE of 0.39 eV (PLS with 30-dim) and 0.41 eV (LDA with 65-dim) when applied to 171 intermediates derived from erythrose, glyceraldehyde, glycerol and propionic acid. These models provide more accurate estimates of adsorption energies than DFTB (1.46 eV), but the lowest MAE of ML methods are still not comparable to DFT. We used Spearman's correlation to show that ML methods are much more reliable for assessing the relative stability of different geometries, providing a more robust route to identifying low-energy structures. Finally, the best aspects of ML and DFTB techniques are combined leading to a new workflow of DFTB+minima hopping $\rightarrow$ DFT $\rightarrow$ ML $\rightarrow$ DFT. This approach allows us to discover 20 new global minima for the 171 adsorbates studied here, which would be missed if only DFTB were used to evaluate candidate structures. Nonetheless, the workflow also has the limitation that there is still uncertainty about whether or not the true global minima is found, since an exhaustive search is not feasible for these complex adsorbates. However, the results indicate that combining physical models and ML predictions is a promising path toward solving this challenging problem. 

\begin{acknowledgement}
CC is supported by a Paper Science and Engineering Fellowship from the Renewable Bioproducts Institute at Georgia Tech. Computational effort was supplied partially by the National Science Foundation under Grant No. MRI-1828187.
\end{acknowledgement}
\bibliography{reference}
\end{document}


\begin{suppinfo}

\subsection{SMILES string generation for adsorbates}
\begin{figure}[h]
    \centering
    \includegraphics[width=0.9\linewidth]{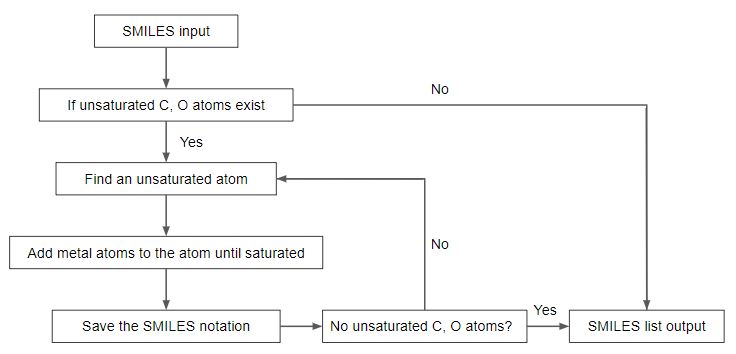}
    \caption{Generating algorithms of SMILES notation of adsorbate with metal atoms}
    \label{fig:algorithm}
\end{figure}

\subsection{Reaction types identified from vector clustering}
\begin{longtable}{|l|l|l|l|l|l|}
\caption{83 clusters from R0 vectors with atomic environment (\# of hydrogen atom and heavy atom surrounding)}
\\ \hline \multicolumn{1}{|c|}{\textbf{cluster}} & \multicolumn{1}{c|}{\textbf{hydrogen\_0}} & \multicolumn{1}{c|}{\textbf{heavy\_atom\_0}} & \multicolumn{1}{c|}{\textbf{hydrogen\_1}} & \multicolumn{1}{c|}{\textbf{heavy\_atom\_1}} & \multicolumn{1}{c|}{\textbf{label}}\\ \hline 
\endfirsthead

\multicolumn{6}{c}%
{{\bfseries \tablename\ \thetable{} -- continued from previous page}} \\
\hline \multicolumn{1}{|c|}{\textbf{cluster}} & \multicolumn{1}{c|}{\textbf{hydrogen\_0}} & \multicolumn{1}{c|}{\textbf{heavy\_atom\_0}} & \multicolumn{1}{c|}{\textbf{hydrogen\_1}} & \multicolumn{1}{c|}{\textbf{heavy\_atom\_1}} & \multicolumn{1}{c|}{\textbf{label}}\\ \hline 
\endhead

\hline \multicolumn{3}{|r|}{{Continued on next page}} \\ \hline
\endfoot

\hline \hline
\endlastfoot
1         & 2                  & 3                     & 1                  & 4                     & C-C       \\
2          & 2                  & 3                     & 0                  & 4                     & C-C       \\
3          & 0                  & 4                     & 0                  & 3                     & C-C       \\
4          & 1                  & 4                     & 1                  & 3                     & C-C       \\
5          & 1                  & 3                     & 0                  & 3                     & C-C  \\
6          & 0                  & 4                     & 1                  & 3                     & C-C       \\
7          & 2                  & 3                     & 1                  & 3                     & C-C       \\
8          & 1                  & 4                     & 0                  & 3                     & C-C       \\
9          & 0                  & 3                     & 2                  & 3                     & C-C       \\
10         & 1                  & 3                     & 1                  & 3                     & C-C  \\
11         & 0                  & 3                     & 0                  & 3                     & C-C   \\
12         & 0                  & 4                     & 0                  & 4                     & C-C       \\
13         & 2                  & 3                     & 2                  & 3                     & C-C       \\
14         & 1                  & 4                     & 0                  & 4                     & C-C       \\
15         & 1                  & 4                     & 1                  & 4                     & C-C       \\
16         & 0                  & 2                     & 1                  & 4                     & C-C       \\
17         & 0                  & 2                     & 1                  & 3                     & C-C   \\
18         & 0                  & 2                     & 2                  & 3                     & C-C       \\
19         & 0                  & 2                     & 0                  & 4                     & C-C \\
20         & 0                  & 2                     & 0                  & 3                     & C-C  \\
21         & 0                  & 2                     & 0                  & 2                     & C-C \\
22         & 1                  & 2                     & 2                  & 2                     & C-C       \\
23         & 2                  & 2                     & 1                  & 4                     & C-C       \\
24         & 2                  & 2                     & 0                  & 3                     & C-C  \\
25         & 0                  & 4                     & 2                  & 2                     & C-C  \\
26         & 2                  & 2                     & 1                  & 3                     & C-C       \\
27         & 2                  & 2                     & 2                  & 3                     & C-C       \\
28         & 0                  & 2                     & 2                  & 2                     & C-C \\
29         & 2                  & 2                     & 2                  & 2                     & C-C       \\
30         & 0                  & 2                     & 1                  & 2                     & C-C \\
31         & 1                  & 2                     & 1                  & 4                     & C-C       \\
32         & 1                  & 2                     & 0                  & 4                     & C-C  \\
33         & 1                  & 2                     & 1                  & 3                     & C-C  \\
34         & 1                  & 2                     & 0                  & 3                     & C-C  \\
35         & 1                  & 2                     & 2                  & 3                     & C-C       \\
36         & 1                  & 2                     & 1                  & 2                     & C-C \\
37         & 3                  & 2                     & 0                  & 4                     & C-C       \\
38         & 3                  & 2                     & 0                  & 3                     & C-C       \\
39         & 3                  & 2                     & 2                  & 3                     & C-C       \\
40         & 1                  & 3                     & 3                  & 2                     & C-C       \\
41         & 3                  & 2                     & 0                  & 2                     & C-C       \\
42         & 2                  & 2                     & 3                  & 2                     & C-C       \\
43         & 1                  & 2                     & 3                  & 2                     & C-C       \\
44         & 1                  & 4                     & 1                  & 2                     & C-O       \\
45         & 1                  & 3                     & 1                  & 2                     & C-O       \\
46         & 0                  & 4                     & 1                  & 2                     & C-O       \\
47         & 0                  & 3                     & 1                  & 2                     & C-O       \\
48         & 2                  & 3                     & 1                  & 2                     & C-O       \\
49         & 2                  & 2                     & 1                  & 2                     & C-O       \\
50         & 0                  & 2                     & 1                  & 2                     & C-O       \\
51         & 2                  & 2                     & 0                  & 2                     & C-O       \\
52         & 1                  & 3                     & 0                  & 2                     & C-O \\
53         & 1                  & 4                     & 0                  & 2                     & C-O       \\
54         & 0                  & 3                     & 0                  & 2                     & C-O \\
55         & 0                  & 4                     & 0                  & 2                     & C-O \\
56         & 2                  & 3                     & 0                  & 2                     & C-O       \\
57         & 1                  & 2                     & 1                  & 2                     & C-O       \\
58         & 1                  & 2                     & 0                  & 2                     & C-O \\
59         & 0                  & 2                     & 0                  & 2                     & C-O \\
60         & 3                  & 1                     & 1                  & 1                     & C-H       \\
61         & 3                  & 2                     & 1                  & 1                     & C-H       \\
62         & 1                  & 1                     & 1                  & 1                     & C-H       \\
63         & 1                  & 4                     & 1                  & 1                     & C-H       \\
64         & 1                  & 3                     & 1                  & 1                     & C-H       \\
65         & 1                  & 2                     & 1                  & 1                     & C-H       \\
66         & 2                  & 2                     & 1                  & 1                     & C-H       \\
67         & 2                  & 3                     & 1                  & 1                     & C-H       \\
68         & 2                  & 1                     & 1                  & 1                     & C-H       \\
69         & 1                  & 2                     & 1                  & 1                     & O-H       \\
70         & 1                  & 1                     & 1                  & 1                     & O-H       \\

71         & 2                  & 3                     &                   &                       & C-M       \\
72         & 1                  & 4                     &                   &                     & C-M       \\
73         & 0                  & 4                     &                   &                     & C-M       \\
74         & 0                  & 5                     &                   &                     & C-M       \\
75         & 1                  & 3                     &                   &                     & C-M       \\
76         & 0                  & 3                     &                   &                     & C-M       \\
77         & 2                  & 2                     &                   &                     & C-M       \\
78         & 3                  & 2                     &                   &                     & C-M       \\
79         & 0                  & 2                     &                   &                     & C-M       \\
80         & 1                  & 2                     &                   &                     & C-M       \\
81         & 0                  & 3                     &                   &                     & O-M       \\
82         & 0                  & 2                     &                   &                     & O-M       \\
83         & 1                  & 2                     &                   &                     & O-M       \\

\hline
\label{tab:cluster_table}
\end{longtable}

\subsection{DFT Details}

A Monkhorst-Pack k-point sampling\cite{monkhorst1976special} of 4$\times$4$\times$1 and a planewave cutoff of 450 eV were used. All surface species were modeled using 3.8034 as lattice constant and vacuum of 10.0 \AA with periodic condition. A BFGS algorithm provided by Atomic Simulation Environment (ASE)\cite{bahn2002object} was applied to the geometry optimization until the maximum force was no more than 0.05 eV/\AA.
The adsorption energy is calculated as follow:
\begin{equation}
    E_{adsorption} = E_{system} - E_{surface} - E_{adsorbate}
    \label{equ:E}
\end{equation}
where $E_{adsorption}$ is the adsorption energy, $E_{system}$ is the total energy of the adsorbate and the Rh slab, $E_{surface}$ is the energy of Rh slab and $E_{adsorbate}$ is the reference energy of adsorbate relative to $CH_{4}$, $H_{2}O$ and $H_{2}$.

\subsection{Confidence Interval Calculation}
The 75\% confidence of the 0.45 eV criterion is calculated as following (assuming $\hat{E_{1}}-\hat{E_{2}}>0.45$ and $\sigma=0.45 eV$):

\begin{gather}
    E - \hat{E} = Z \sim Normal(0, \sigma^2) \\
    E_{1} = \hat{E_{1}}+Z \sim Normal(\hat{E_{1}},\sigma^2) \\
    E_{2} = \hat{E_{2}}+Z \sim Normal(\hat{E_{2}}, \sigma^2) \\
    E_{1} \independent E_{2} \Rightarrow E_{1}-E_{2} \sim Normal(\hat{E_{1}}-\hat{E_{2}}, 2\sigma^2) \\
    P(E_{1}-E_{2}<0) = \Phi(\frac{\hat{E_{2}}-\hat{E_{1}}}{2\sigma^2}) = 1-\Phi(\frac{\hat{E_{1}}-\hat{E_{2}}}{2\sigma^2})\\
    \frac{\hat{E_{1}}-\hat{E_{2}}}{\sqrt{2}\sigma} > \frac{1}{\sqrt{2}} \Rightarrow P(E_{1}-E_{2}<0)<1-\Phi(\frac{1}{\sqrt{2}})=0.25\\
    P(E_{1}<E_{2}|\hat{E_{1}-\hat{E_{2}}}>0.45)<0.25 \\
    \Rightarrow P(E_{1}>E_{2}|\hat{E_{1}-\hat{E_{2}}}>0.45)>0.75 
\end{gather}

The following files are available free of charge. (\url{https://github.com/cchang373/Rh_paper})
\begin{itemize}
  \item regression: contains OLS, PLS and LDA regression scripts and the DFT regression data in .json file
  \item all\_adsorbates.smi: contains SMILES notation of all adsorbates used in this study
  \item model: contains the Mol2Vec training data (groups.smi) and the Mol2Vec corpus data (groups.cp)
  \item spearman: contains Spearman's correlation coefficient calculation script and the DFT, PLS and LDA regression data in .json file
  \item dft\_diff.json: contains SMILES notations of previous lowest energy geometry, new lowest energy geometry and energy difference for the adsorbates found by ML models 
\end{itemize}

\end{suppinfo}
\bibliography{reference}